# Will recent advances in AI result in a paradigm shift in Astrobiology and SETI?

Joe Gale[1], Amri Wandel[2]
The Hebrew University of Jerusalem and
Hugh Hill[3]
International Space University (ISU)

[1]gale.joe@mail.huji.ac.il  [2]amri@mail.huji.ac.il  [3]hugh.hill@isunet.edu

[1]Institute of Life Sciences.
[2]Racah Institute of Physics.
[3]Strasbourg Central Campus.

**Abstract**

The steady advances in computer performance and in programming raise the concern that the ability of computers would overtake that of the human brain, an occurrence termed "the Singularity". While comparing the size of the human brain and the advance in computer capacity, the Singularity has been estimated to occur within a few decades although the capacity of *conventional* computers may reach its limits in the near future.

However, in the last few years, there have been rapid advances in Artificial Intelligence (AI). There are already programs that carry out pattern recognition and self-learning which, at least in limited fields such as chess and other games, are superior to the best human players. Furthermore, the quantum computing revolution, which is expected to vastly increase computer capacities, is already on our doorstep.

It now seems inevitable that the Singularity will arrive within the foreseeable future. Biological life, on Earth and on extraterrestrial planets and their satellites, may continue as before, but humanity could be 'replaced' by computers. Older and more advanced intelligent life forms, possibly evolved elsewhere in the universe, may have passed their Singularity a long time ago. Post Singularity life would probably be based not on biochemical reactions but on electronics. Their communication may use effects such as quantum entanglement and be undetectable to us. This may explain the Fermi paradox or at least the "Big Silence" problem in SETI.

**Introduction**

Recent major breakthroughs in computing and Artificial Intelligence (AI) may change human existence on Earth and our thinking in astrobiology, especially in relation to the Search for Extraterrestrial Intelligence (SETI).

Estimates of the values of the terms in the well-known Drake formula, which puts together the chances for contact with intelligent life forms, have made great progress in the last decade. The Kepler mission provided estimates of the values of three factors in the equation, which utill the Kepler mission were uncertain: the fraction of stars with planets, the average number of planets per star in the habitable zone and the fraction of Earth-size planets. These parameters, are now believed to be of order unity (e.g. Batalha *et al*. 2013; Dressing and Charbonneau, 2015; Wandel 2017).

Our main ignorance remains in the last three terms: (i) the chances for the evolution of biological life; (ii) the probability of intelligence; and (iii) the lifespan of a technological communicating civilization. As may be inferred from the evolution of life on Earth, planets with primitive biological life may be quite abundant (Wandel 2015; Gale & Wandel 2017). However, if the last two terms were small, intelligent and communicating civilizations could be exceedingly rare (Wandel 2017).

Complex life is a relatively recent development on Earth, where, for approximately 3.5Gy (about a quarter of the lifetime of the universe) only simple mono-cellular life appeared (Fig. 1). Life elsewhere may have evolved at a different rate. It is reasonable to assume that many instances of any intelligent life would be more advanced than humanity.

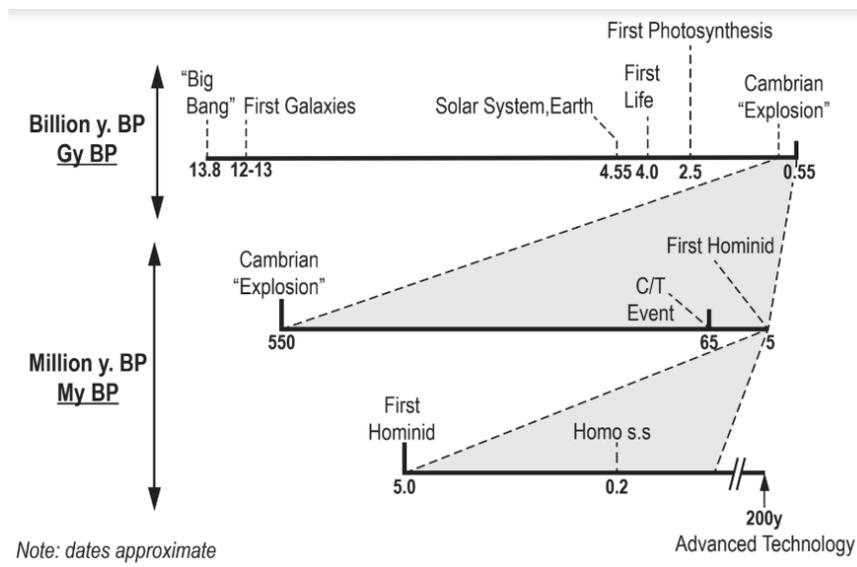

Fig. 1. Time line for the evolution of life on Earth. Data from Lyon et al, 2014 and elsewhere

Science fiction is replete with humanoids, despite their short appearance on Earth. Homo sapiens has only been around for about 200,000 years, and advanced technology for only two centuries (Fig.1). Science fiction authors have described advanced and even intelligent computers, nearly always stopping at the inability of computers to carry out advanced pattern recognition, the basis of intelligent thinking. For this, a huge computing ability is considered essential, as in the human brain.

**Moore's Law**

In the last decades many scientists, futurists and Science Fiction authors have proposed that in the near future computers would have the same calculating capacity as human brains, which have ~$10^{11}$ neurons, each connected to thousands of synapses. This proposition has been based on projections of "Moore's law" of the evolution of computers. This is an observation that the number of transistors in integrated circuits doubles about every two years (Moore 1965). Remarkably, that prediction has held till today, fifty years later (Fig. 2). This advance has probably reached its almost inevitable limit, as a result of the nano-meter size of today's transistors and the impossibility of providing electrical insulation at this scale (Hennessey and Patterson 2019). However, with the development of quantum computers, even this limit appears to be passable.

Figure 2. Moore's Law: the number of transistors per microprocessor vs. time

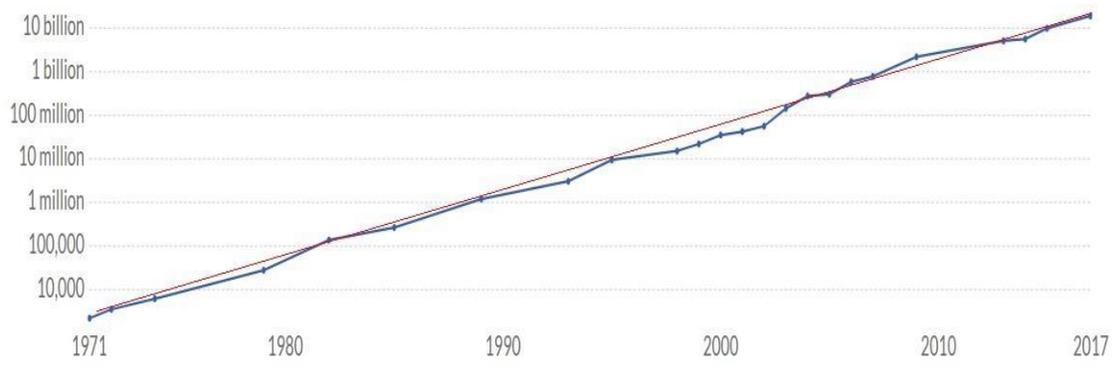

Data from Rupp (2017). The straight line shows an exponential growth, doubling every two years.

Vernor Vinge, a mathematician and science fiction author, introduced the term "Singularity" to describe the convergence of the capacities of the human brain and computers (Vinge 1993). This term was previously used in mathematics and natural sciences to describe a point where some quantity becomes infinite, or ill behaved. In the context of the advance of technology, the concept received support from eminent polymath John von Neumann, quoted as saying that "the singularity will be reached when technological progress becomes incomprehensively rapid and complicated" (Shanahan 2015). The "Singularity" concept[1] and its possible consequences were popularized by the futurist author Ray Kurzweil in his book "The Singularity is near" (Kurzweil 2005). Extrapolating Moore's "Law", he estimated that this event would occur in about 2050. Others have suggested that it may come even earlier if quantum computers are realized [2].

**Brains**

Neurologists have long pointed out the poor correlation between brain size and advanced thought. Some collated data on the relative sizes of different animal brains (here, the number of neurons) are shown in Table 1 (data collated from Herculano-Houzel 2009 and elsewhere).

Table 1. Some animal brain sizes

|   | Number of Neurons (in Millions) |
|---|---|
| Sperm Whales | - 250,000 |
| African Elephants | - 207,000 |
| Homo sapiens sapiens | - 86,000 |
| Gorillas | - 33,400 |
| Ratus ratus | - 103 |
| Bees (Anthophila) | - 1 |

As apparent from the number of neurons (Table 1), brain size is not well correlated to intelligence. For example, humans have less than half the brain size of the not very intelligent elephants. Of particular significance is the very small brain capacity of insects; here e.g., bees. With just 1 million neurons, bees carry out: Flight; Navigation; Hunting and Gathering; Communication with other bees; Social organization; inheritable home Architecture; preparation of hive Emergency Procedures, etc. etc. (for numerical ability in bees, see Howard et al 2018). Clearly, programming seems to be no less important than brain capacity.

---

[1] https://en.wikipedia.org/wiki/Technological_singularity retrieved 30.8.2019
[2] https//singularityhub.com/2019/02/26/quantum-computing-now-and-in-the-not-too-distant-future retrieved 31.8.201

Human brains have developed an extraordinary ability for pattern recognition, which has long been thought to be beyond the ability of computers.

**Rapid Advances in Computers**

Recently, computers have been catching up. In 1997, IBM's program Deep Blue beat reigning chess champion Gary Kasparov. However, it was mainly a victory for brute force and the use of a small number of preset strategies [3]. In 2016, the program AlphaGO, fortified with neural networks and learning modes, defeated Lee Sedol, the Go world champion[4].

Recently Google described "Alpha Zero", a self- learning, pattern- recognizing, AI program (Silver et al 2018). Given just the basic rules of Chess, Shogi and Go, this program plays itself millions of times over, selecting and remembering the most favorable winning strategies. The program has beaten human masters in all three games, by using strategies *previously unknown to programmers.*

The Google program is predicted to be able to solve many hitherto almost intractable problems. For example, in Astronomy - searching data banks for enigmatic radio bursts, in Medicine - reviewing millions of combinations of illness-causing gene and intermediate interactions, as opposed to single gene errors, and in Meteorology - predicting weather patterns from huge data banks.

Brown and Sandholm (2019) have described an AI program, based on Neumann and Nash game theory, which challenges hitherto unbeatable Multi-player poker games.

For many years, the concept of quantum-based computers has been predicted as a way forward, once the physical limits of transistors are reached. In theory, a quantum-based computer would be able to utilize an analog characteristic of multiple positions between zero and one, a limitation of today's digital technology (Osorio 2000). Until recently, intrinsic difficulties caused by de-coherence, an instability due to the statistical nature of quantum phenomena, and the sensitivity of such computers to vibration, temperature fluctuations and external electromagnetic waves, have prevented practical realization. However, a combination of improved hardware and error reducing software has very recently promised to overcome these problems (Niu et al 2019; Levine Y. et al 2019). It has been predicted that Quantum Supremacy, a term used to describe the point at which quantum computers will out-perform the largest digital computers, could be reached by the end of 2019 (Hartnett 2019).

---

[3] https//en.wikipedia.org/wiki/deep_blue_versus_gary_kasparov, retrieved 30.8.2019.

[4] https://en.wikipedia.org/wiki/AlphaGo_versus_Lee_Sedol , retrieved 30.8.2019.

These advances in computer capacity and AI suggest that the Singularity may indeed be imminent. It may not come in the very near future but it will arrive, eventually. When it does, Homo sapiens, sapiens may move from the bio-chemical realm to the electronic, or to some hybrid combination, with hardly predictable implications for humans (Kurzweil 2005).

While life appeared on Earth some 4Gy ago, for 3.5Gy only mono-cellular life existed. Complex life evolved only some 0.5Gy ago, when a high oxygen atmosphere developed. Homo s.s. appeared just 200,000y ago, and advanced technology - only in the last two centuries (Fig.1).

**Consequences of the Singularity for Astrobiology**

The search for Extraterrestrial biological life will probably continue along the present lines. Singularity only applies to advanced civilizations, and consequently to SETI. Since some putative extra-terrestrial civilizations have probably had more time than us to evolve, they might have long passed their Singularity.

Once the Singularity is attained, optimal conditions for advanced, complex life may fundamentally change. We may find that the universe is populated by autonomous, intelligent, space ships driven by quantum computing devices. If future computers are anything like ours of today (of course they may be quite different), they may prefer dry, non-corrosive conditions, with low temperatures, which would provide a "quiet" electronic environment and enable super-conductivity (Dougherty & Kimel 2012). They would preferably operate not from the surface of "habitable" planets but in space, perhaps parked in dynamically stable Lagrangian points. However, because of the huge interstellar distances, SETI would not be able to distinguish if broadcasts originated from a planet or rather from a Lagrangian point of the planet and its moon or sun.

If an alien civilization ever visited Earth long ago, it may have parked its beacons in one of the Earth-Moon or Sun-Earth Lagrangian points. An attempt to find such artifacts, by photography was unsuccessful (Freitas Jr. and Valdes 1980).

Furthermore, post-singularity civilizations may not be communicating by electromagnetic waves but rather by quantum entanglement, which is only in its very first stages of development on Earth (Ursin et al 2007; Kumar et al 2019). This may provide yet another solution to the Fermi paradox, or at least to the Big Silence problem in SETI.

As for humanity's future, perhaps the superior brains of the new computers will solve Earth and humanities many problems.  Science Fiction is replete with predictions, some reasonable, some disturbing, such as that mortality can be overcome by downloading the brain's contents to computers. In 2014, Stephen Hawkins stated in an interview[5] "The development of full artificial intelligence could spell the end of the human race." In other words, AI and the Singularity may be humanity's greatest and last advance.

---

[5] https://www.bbc.com/news/technology-30290540 retrieved 30.8.2019